\documentclass[conference,a4paper]{APSIPA2021}
\usepackage{multirow}
\usepackage{amsmath}
\usepackage[psamsfonts]{amssymb}
\usepackage{amsxtra}
\usepackage{threeparttable}
\usepackage{algorithm}
\usepackage{graphicx}
\usepackage{subfigure}
\usepackage{algorithm}
\usepackage{algpseudocode}
\usepackage{multirow}
\usepackage{xcolor,cancel}
\usepackage[normalem]{ulem}
\usepackage{cite}

\begin{document}

\title{A Study on Speech Enhancement Based on Diffusion Probabilistic Model}

\author{%
\authorblockN{%
Yen-Ju Lu\authorrefmark{1},
Yu Tsao\authorrefmark{1} and
Shinji Watanabe\authorrefmark{2}
}
\authorblockA{%
\authorrefmark{1}
Research Center for Information Technology Innovation, Academia Sinica, Taipei, Taiwan \\
E-mail: \{neil.lu, yu.tsao\}@citi.sinica.edu.tw}
\authorblockA{%
\authorrefmark{2}
Language Technology Institute, Carnegie Mellon University, Pittsburgh, PA, United States \\
E-mail: shinjiw@cmu.edu}
}

\maketitle
\thispagestyle{empty}

\begin{abstract}

Diffusion probabilistic models have demonstrated an outstanding capability to model natural images and raw audio waveforms through a paired diffusion and reverse processes. The unique property of the reverse process (namely, eliminating non-target signals from the Gaussian noise and noisy signals) could be utilized to restore clean signals. Based on this property, we propose a diffusion probabilistic model-based speech enhancement (DiffuSE) model that aims to recover clean speech signals from noisy signals. The fundamental architecture of the proposed DiffuSE model is similar to that of DiffWave--a high-quality audio waveform generation model that has a relatively low computational cost and footprint. To attain better enhancement performance, we designed an advanced reverse process, termed the supportive reverse process, which adds noisy speech in each time-step to the predicted speech. The experimental results show that DiffuSE yields performance that is comparable to related audio generative models on the standardized Voice Bank corpus SE task. Moreover, relative to the generally suggested full sampling schedule, the proposed supportive reverse process especially improved the fast sampling, taking few steps to yield better enhancement results over the conventional full step inference process. 

\end{abstract}

\section{Introduction}
The goal of speech enhancement (SE) is to improve the intelligibility and quality of speech, by mapping distorted speech signals to clean signals. The SE unit has been widely used as a front-end processor in various speech-related applications, such as speech recognition \cite{li2014overview,erdogan2015phase,chen2015speech}, speaker recognition \cite{michelsanti2017conditional}, assistive hearing technologies \cite{healy2019optimal, lai2016deep}, and audio attack protection \cite{yang2020characterizing}. Recently, deep neural network (DNN) models have been widely used as fundamental tools in SE systems, yielding promising results \cite{lu2013speech,wang2014training,xia2014wiener,xu2014regression, siniscalchi2021vector,qi2020exploring,le2013ensemble}. Compared to traditional SE methods, DNN-based methods can more effectively characterize nonlinear mapping between noisy and clean signals, particularly under extremely low signal-to-noise (SNR) scenarios and/or non-stationary noise environments \cite{tan2019real,kolbaek2016speech,qi2019theory}. 

Traditional SE methods calculates the noisy-clean mapping through the discriminative methods in  \textit{time-frequency (T-F) domain} or \textit{time domain}. For the T-F domain methods, the time-domain speech signals are first converted into spectral features through a short-time Fourier transform (STFT). The mapping function of noisy to clean spectral features is then formulated by a direct mapping function \cite{lu2013speech, xu2014regression}, or a masking function \cite{wang2014training,weninger2015speech,subramanian2018student}. The enhanced spectral features are reconstructed to time-domain waveforms with the phase of the noisy speech based on the inverse STFT operation \cite{wang2018supervised}. As compared with T-F domain methods, it has been shown that the time-domain SE methods can avoid the distortion caused by inaccurate phase information \cite{fu2018end, germain2018speech}. 
To date, several audio generation models have been directly applied to or moderately modified to perform SE, estimating the distribution of the clean speech signal, such as generative adversarial networks (GAN) \cite{pascual2017segan,soni2018time,fu2019metricgan}, autoregressive models \cite{qian2017speech}, variational autoencoders (VAE) \cite{leglaive2020recurrent}, and flow-based models \cite{strauss2021flow}.  

The diffusion probabilistic model, proposed in \cite{sohl2015deep}, has shown strong generation capability. The diffusion probabilistic model includes a diffusion/forward process and a reverse process. The diffusion process converts clean input data to an isotropic Gaussian distribution by adding Gaussian noise to the original signal at each step. In the reverse process, the diffusion probabilistic model predicts a noise signal and subtracts the predicted noise signal from the noisy input to retrieve the clean signal. The model is trained by optimizing the evidence lower bound (ELBO) during the diffusion process. Recently, the diffusion probabilistic models have been shown to provide outstanding performance in generative modeling for natural images \cite{ho2020denoising,nichol2021improved}, and raw audio waveforms \cite{kong2020diffwave,liu2021diffsvc}. As reported in \cite{kong2020diffwave}, the DiffWave model, formed by the diffusion probabilistic model, can yield state-of-the-art performance on either conditional or unconditional waveform generation tasks with a small number of parameters.

In this study, we propose a novel diffusion probabilistic model-based SE method, called DiffuSE. The basic model structure of DiffuSE is similar to that of Diffwave. Since the target task is SE, DiffuSE uses the noisy spectral features as the conditioner, rather than the clean Mel-spectral features used in DiffWave. Meanwhile, different from the derived equation of the diffusion model, we combine the noisy speech signal into the reverse process instead of the isotropic Gaussian noise.  To further improve the quality of the enhanced speech, we pretrained the model using clean Mel-spectral features as a conditioner. After pretraining, we replaced the conditioner with noisy spectral features, reset the parameters in the conditioner encoder, and preserved other parameters for the SE training.

The contributions of this study are three-fold: (1) It is the first study to apply the diffusion probabilistic model to the SE tasks. (2) We propose a novel supportive reverse process, specifically for the SE task, which combines the noisy speech signals during the reverse process. (3) The experimental results confirm the effectiveness of DiffuSE, which provides comparable or even better performacne as compared to related time-domain generative SE methods.

The remainder of this paper is organized as follows. We present the diffusion models in Section \uppercase\expandafter{\romannumeral2} and introduce the DiffuSE architecture in Section \uppercase\expandafter{\romannumeral3}. We provide the experimental setting in Section \uppercase\expandafter{\romannumeral4}, report the results in Section \uppercase\expandafter{\romannumeral5}, and conclude the paper in Section \uppercase\expandafter{\romannumeral6}.

\section{Diffusion probabilistic models}
\label{sec:format}

This section introduces the diffusion and the reverse procedures of the diffusion probabilistic model. A detailed mathematical proof of the model's ELBO can be found in \cite{ho2020denoising}, and we only discuss the diffusion and reverse processes with their algorithm in this section.


\begin{figure}[h]
 \centering
 \includegraphics[width=\linewidth]{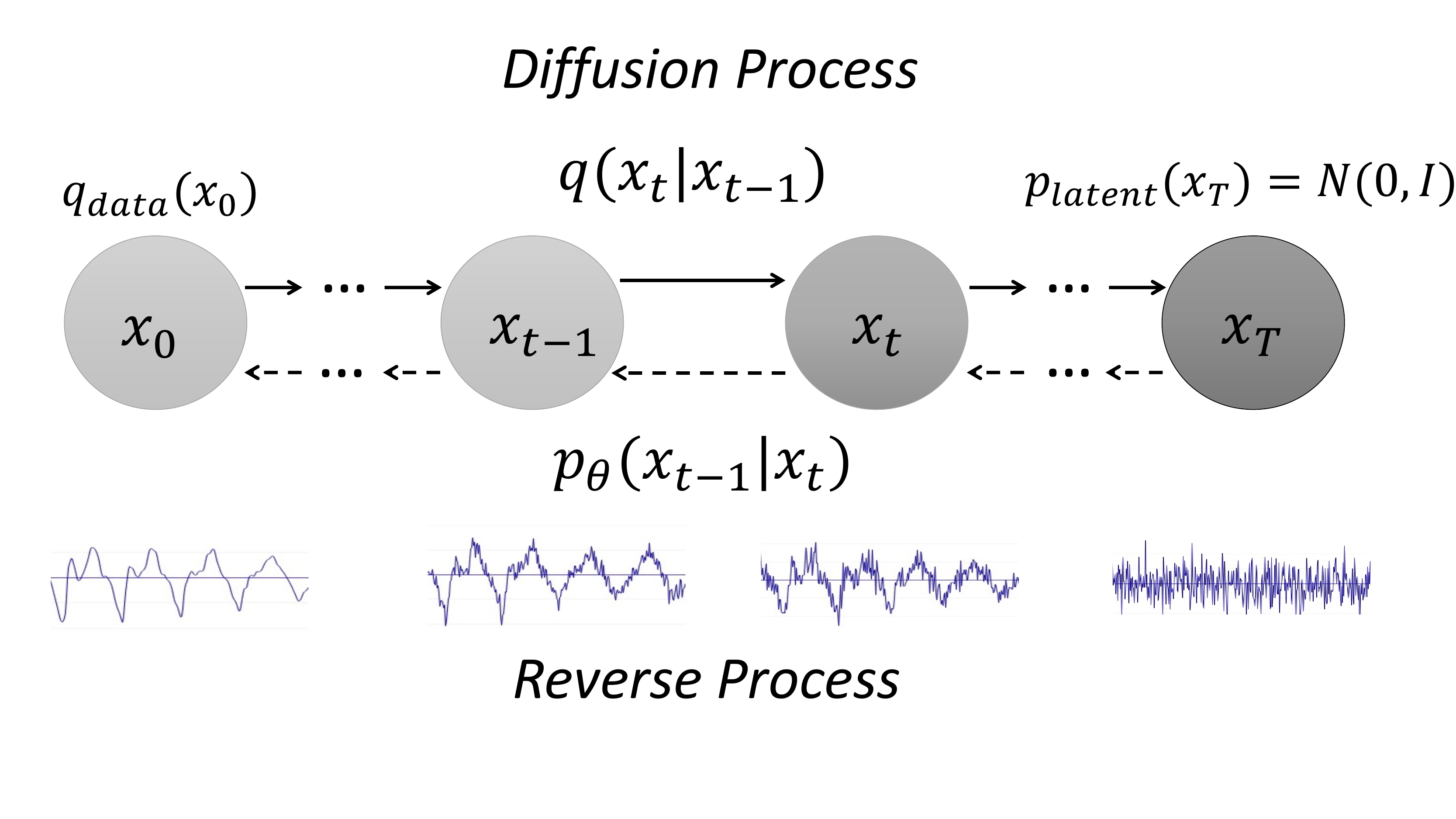}
 \caption{The diffusion process (solid arrows) and reverse processes (dashed arrows) of the diffusion probabilistic model.} 
 \label{fig:DPM}
\end{figure}

\begin{algorithm}[tp]
  \caption{Training}\label{alg:training}
  \begin{algorithmic}
      \For{$i=1,2,\cdots,N_{iter}$}
        \State {Sample $x_0{\sim}q_{\text{data}}, \epsilon{\sim}N(0,I),$ and} 
        \State {$t{\sim}Uniform(\{1,\cdots,T\})$}
        \State {Take gradient step on}
        \State {$\nabla_\theta \parallel \epsilon-\epsilon_{\theta}(\sqrt{\bar{\alpha}_t}x_0 + \sqrt{1-\bar{\alpha}_t}\epsilon,t)\parallel^2_2$}
        \State {according to Eq. \ref{train eq}}
      \EndFor
  \end{algorithmic}
\end{algorithm}

\subsection{Diffusion and Reverse Processes}
\label{subsec:Diff and Rev Process}
A diffusion model of $T$ steps is composed of two processes: the \textit{diffusion} process with steps $t = (0,1,\cdots,T)$ and the \textit{reverse} process $t = (T,T-1,\cdots,0)$ \cite{sohl2015deep}. The input data distribution of the diffusion process is defined as $q_{\text{data}}(x_0)$ on $\mathbb{R}^L$, where $L$ is the data dimension. 
$x_t\in \mathbb{R}^L$ is a step-dependent variable at diffusion step $t$ with the same dimension $L$. 
The diffusion and the reverse processes are illustrated in Figure \ref{fig:DPM}.

In Figure \ref{fig:DPM}, The solid arrows are the diffusion process from data $x_0$ to the latent variable $x_T$, represented as:
\begin{align}
  q(x_1,\cdots,x_T|x_0) 
  & = \prod_{t=1}^{T} q(x_t|x_{t-1}),
  \label{diffuse eq1}
\end{align}
where $q(x_t|x_{t-1})$ is formulated by a fixed Markov chain, $N(x_t;\sqrt{1-\beta_{t}}x_{t-1},\beta_{t}I)$, with a small positive constant ratio $\beta_{t}$, and the Gaussian noise is added to the previous distribution $x_{t-1}$. The overall process gradually converts data $x_0$ to a latent variable with an isotropic Gaussian distribution of $p_{\text{latent}}(x_T) = N(0,I)$, according to the predefined schedule $\beta_{1},\cdots,\beta_{T}$. 

The sampling distribution at the $t$-th step, $x_t$, can also be derived from the distribution of $x_0$ in a closed form by marginalizing $x_1, \dots, x_{t-1}$ as:
\begin{equation}
  q(x_t|x_0) = N(x_t;\sqrt{\bar{\alpha}_t}x_0,(1-\bar{\alpha}_t)I),
  \label{diffuse eq2}
\end{equation}
where $\alpha_t = 1-\beta_t$ and $\bar{\alpha}_t =  \prod_{s=1}^{t} \alpha_s$. Empirically, we can sample the $t$-th step distribution $x_{t}$ from the initial data $x_{0}$ directly. In contrast, The dashed arrows in Figure \ref{fig:DPM} are the reverse process, converting the latent variable $x_{T}$ to $x_{0}$, which is also defined by a Markov chain:
\begin{equation}
  p_\theta(x_0,\cdots,x_{T-1}|x_T) = \prod_{t=1}^{T} p_\theta(x_{t-1}|x_t),
  \label{reverse eq1}
\end{equation}
where $p_\theta(\cdot)$ is the distribution of the reverse process with learnable parameter $\theta$.
Because the marginal likelihood $p_{\theta}(x_0) = \int p_\theta(x_0,$ $\cdots,$ $x_{T-1}|x_T)\cdot p_{\text{latent}}(x_T) dx_{1:T}$ is intractable for calculations in general, the model should be trained using ELBO. Recently, \cite{ho2020denoising} showed that under a certain parameterization, the ELBO could be calculated using a closed-form solution.
 
\begin{algorithm}[tp]
  \caption{Sampling}\label{alg:sampling}
  \begin{algorithmic}
      \State {Sample $x_T{\sim}p_{\text{latent}} =N(0,I),$} 
      \For{$t=T,T-1,\cdots,1$}
        \State {Compute $\epsilon_{\theta}(x_t,t)$ and ${\sigma_t}$}
        \State {Sample $x_{t-1}\sim p_{\theta}(x_{t-1}|x_t)=$}
        \State {$N(x_{t-1};\frac{1}{\sqrt{\alpha_t}}(x_t-\frac{\beta_t}{\sqrt{1-\bar{\alpha}_t}}\epsilon_\theta(x_t,t)), \sigma^2_tI)$}
        \State {according to Eq. \ref{sample eq}}
      \EndFor
      \State \textbf{return} $x_0$
  \end{algorithmic}
\end{algorithm}

\subsection{Training through Parameterization}
\subsubsection{Parameterization}
The transition probability in the reverse process $p_{\theta}(x_{t-1}|x_t)$ in Eq.~\ref{reverse eq1} can be represented by two parameters, $\mu_\theta$ and $\sigma_\theta$, as $N(x_{t-1};$ $\mu_\theta(x_t,t),$ ${\sigma_\theta(x_t,t)}^2I)$, with a learnable parameter $\theta$. $\mu_\theta$ is an $L$-dimensional vector, that estimates the mean of the distribution of $x_{t-1}$. $\sigma_\theta$ denotes the standard deviation (a real number) of the $x_{t-1}$ distribution. Note that both values take two inputs: the diffusion step $t\in \mathbb{N}$, and variable $x_t\in \mathbb{R}^L$.
Further, Eq. \ref{diffuse eq2} can also be reparameterized as $x_t(x_0,\epsilon) = \sqrt{\bar{\alpha}_t}x_0 + \sqrt{1-\bar{\alpha}_t}\epsilon$ for $\epsilon \sim N(0,I)$. $\sigma_\theta(x_t,t)$ was set to $\sigma_t$ as a time-dependent parameter. 

\subsubsection{Training and Sampling}
In the reverse process, $p_{\theta}(x_{t-1}|x_t)$ in Eq.~\ref{reverse eq1} aims to predict the previous distribution by the current mixed data with extra Gaussian noise added in the diffusion process. Therefore, the predicted mean $\mu_\theta$ is estimated by eliminating the Gaussian noise $\epsilon$ in the mixed data $x_t$. According to the derivations in \cite{ho2020denoising}, $\mu_\theta$ can be predicted by a given $x_t$ and $t$ as Eq. \ref{mu eq}:

\begin{equation}
 \mu_\theta(x_t,t) =
 \frac{1}{\sqrt{\alpha_t}}(x_t-\frac{\beta_t}{\sqrt{1-\bar{\alpha}_t}}\epsilon_\theta(x_t,t)), 
 \label{mu eq}
\end{equation}

Note that the real Gaussian noise added in the diffusion process $\epsilon$ is unknown in the reverse process. Therefore, the model $\epsilon_\theta$ should be designed to predict $\epsilon$. In contrast, $\sigma_t$, the standard deviation of the $x_{t-1}$, can be fixed to a constant for every step t as Eq \ref{sigma eq}:

\begin{equation}
 \sigma_t =\widetilde{\beta}_t^{\frac{1}{2}}, \textnormal{where } \widetilde{\beta}_t=
 \left\{
     \begin{array}{lr}
     \frac{1-\bar{\alpha}_{t-1}}{1-\bar{\alpha}_t}\beta_t \textnormal{ for } t>1, \\
     \beta_0 \textnormal{ for } t=0,
     \end{array}
\right.
 \label{sigma eq}
\end{equation} 

Therefore, for predicting $\mu_\theta(x_t,t)$ in the reverse process, the model parameters $\theta$ aim to estimate the Gaussian noise $\epsilon_\theta(x_t,t)$ by input $x_t$ and $t$. During the diffusion process, the training loss of the model is defined to reduce the distance of the estimated noise $\epsilon_\theta(x_t,t)$ and the Gaussian noise $\epsilon$ in the mixed data $x_t$, as shown in Eq. \ref{train eq}.

\begin{equation}
\nabla_\theta \parallel \epsilon-\epsilon_{\theta}(\sqrt{\bar{\alpha}_t}x_0 + \sqrt{1-\bar{\alpha}_t}\epsilon,t)\parallel^2_2
  \label{train eq}
\end{equation}

After the training process,  $x_{t-1}$ was computed using Eq. \ref{sample eq} where $z \sim N(0,I)$. 
 
\begin{equation}
  x_{t-1} = \frac{1}{\sqrt{\alpha_t}}\left(x_t-\frac{\beta_t}{\sqrt{1-\bar{\alpha}_t}}\epsilon_\theta(x_t,t)\right)+ \sigma_tz,
  \label{sample eq}
\end{equation}

To summarize, the model is trained during the diffusion process by estimating the Gaussian noise $\epsilon$ inside the mixed-signal $x_t$, and samples the data $x_0$ through the reverse process. We describe the diffusion and reverse processes in Algorithms \ref{alg:training} and \ref{alg:sampling}, respectively. Table \ref{tab:parameters} lists the parameters of the diffusion probabilistic models.

\begin{table}[hp]
\centering
\caption{Parameters in the diffusion probabilistic models}
\label{tab:parameters}
\begin{tabular}{lll}
\hline
\multicolumn{1}{|l|}{Process} & \multicolumn{1}{l|}{Parameter} & \multicolumn{1}{l|}{Meaning} \\ \hline
\multicolumn{1}{|l|}{\multirow{4}{*}{\begin{tabular}[c]{@{}l@{}}Diffusion\\ Process\end{tabular}}} & \multicolumn{1}{l|}{$\alpha_t$} & \multicolumn{1}{l|}{ratio of $x_{t-1}$ in $x_t$} \\ \cline{2-3} 
\multicolumn{1}{|l|}{} & \multicolumn{1}{l|}{$\beta_t$} & \multicolumn{1}{l|}{ratio of noise added in $x_t$} \\ \cline{2-3} 
\multicolumn{1}{|l|}{} & \multicolumn{1}{l|}{$\bar{\alpha}_t$} & \multicolumn{1}{l|}{ratio of $x_0$ in $x_t$} \\ \cline{2-3} 
\multicolumn{1}{|l|}{} & \multicolumn{1}{l|}{$\epsilon$} & \multicolumn{1}{l|}{isotropic Gaussian noise} \\ \hline
\multicolumn{1}{|l|}{\multirow{3}{*}{\begin{tabular}[c]{@{}l@{}}Reverse\\ Process\end{tabular}}} & \multicolumn{1}{l|}{$\epsilon_\theta $} & \multicolumn{1}{l|}{ predicted noise from model $\theta$} \\ \cline{2-3} 
\multicolumn{1}{|l|}{} & \multicolumn{1}{l|}{$\mu_\theta$} & \multicolumn{1}{l|}{predicted mean from model $\theta$} \\ \cline{2-3} 
\multicolumn{1}{|l|}{} & \multicolumn{1}{l|}{$\sigma_t$} & \multicolumn{1}{l|}{standard deviation} \\ \hline
 &  &  \\
 &  & 
\end{tabular}
\end{table}


\section{DiffuSE architecture}
\label{sec:pagestyle}

In the proposed DiffuSE model, we derive a novel supportive reverse process to replace the original reverse process, to eliminate noise signals from the noisy input more effectively. 

\subsection{Supportive Reverse Process}
In the original diffusion probabilistic model, the Gaussian noise is applied in the reverse process. Since the clean speech signal was unseen during the reverse process, the calculated speech signal, $x_t$, may be distorted during the reverse process from step $T,\cdots,t+1$. To address this issue, we proposed a supportive reserve process, starting the sampling process from the noisy speech signal $y$, and combining $y$ at each reverse step while reducing the additional Gaussian signal. 

The noisy speech signal $y \in \mathbb{R}^{L} $ can be considered as a combination of the clean speech signal $x_0$ and background noise $n \in \mathbb{R}^{L}$, as $y = x_0 + n$. In the supportive reserve process, we define a new valuable $\hat{\mu}_\theta(x_t,t)$, which is a combination of noisy speech $y$ and the predicted $\mu_\theta(x_t,t)$ as shown in Eq. \ref{supportive reverse eq1}:

\begin{equation}
  \hat{\mu}_\theta(x_t,t)  = (1-\gamma_t)\mu_\theta(x_t,t) +  \gamma_t\sqrt{\bar{\alpha}_{t-1}} y
  \label{supportive reverse eq1}
\end{equation}
where $\hat{\mu}_\theta(x_t,t)$ can be formulated as $\hat{\mu}_\theta(x_t,t)  = \sqrt{\bar{\alpha}_{t-1}}(x_0 + \gamma_tn)$ from the mean of $x_{t-1}$ is known as $\sqrt{\bar{\alpha}_{t-1}}x_0$ in the diffusion process. Therefore, we filled the remaining part of noise by the Gaussian signal with the independent assumption as Eq. \ref{supportive reverse eq2}:

\begin{equation}
  \hat{\sigma}_t  = \sqrt{\sigma_t^2 - \gamma_t^2\bar{\alpha}_{t-1}}
  \label{supportive reverse eq2}
\end{equation}

In diffusion models, $\epsilon_\theta(x_t,t)$ is used to predict the noise signal $\epsilon$ from $x_t = \sqrt{\bar{\alpha}_t}x_0 + \sqrt{1-\bar{\alpha}_t}\epsilon$. For the SE task, instead of following the original reverse equations derived from the diffusion process, the objective of $\epsilon_\theta(x_t,t)$ could also be considered as predicting the non-speech part $\epsilon$, which is then used to recover the clean speech signal $x_0$ from the mixed-signal $x_t$.
Therefore, although the supportive reverse process replaces the combination of predicted mean and Gaussian noise by the noisy signal,  $\epsilon_\theta$ still has the ability to predict the non-speech components from the noisy signal $x_t$ at the $t$-th step based on the learned knowledge about different speech-noise combinations during the diffusion process. In addition, because $x_t$ is a combination of the clean speech signal $x_0$ and the Gaussian noise $\epsilon$, to reach a more efficient clean speech recovery, the supportive reverse process directly uses the noisy speech signal $y$ as the input of the reverse process rather than the Gaussian noise. Meanwhile, at each reverse step, the supportive reverse process combines $\mu_\theta(x_t,t)$ with the noisy speech $y$ and the Gaussian noise $z$ to form the input $x_t$ of $\epsilon_\theta(x_t,t)$. After the overall reverse process is completed, we follow the suggestion in \cite{abd2008speech,defossez2020real} to combine the enhanced and original noisy signal to obtain the final enhanced speech. 
The detailed procedure of the supportive reverse process is shown in Algorithm \ref{alg:supportive reverse process}.

\begin{algorithm}[tp]
  \caption{Supportive Reverse Sampling}\label{alg:supportive reverse process}
  \begin{algorithmic}
      \State {$x_T = y,$} 
      \For{$t=T,T-1,\cdots,1$}
        \State {Compute $\hat{\mu}_\theta(x_t,t)$ and ${\sigma_t}$ }
        \State {Sample $z \sim N(0,I)$ if $t>1, $ else $ z=0$}
        \State {$x_{t-1} = \hat{\mu}_\theta(x_t,t)+ \sqrt{\sigma_t^2 - \gamma_t^2\bar{\alpha}_{t-1}}z$}
        \State {according to Eq. \ref{supportive reverse eq1} and \ref{supportive reverse eq2}}
      \EndFor
      \State \textbf{return} $x_0$
  \end{algorithmic}
\end{algorithm}




\subsection{Model Structure}

\subsubsection{DiffWave Architecture}
The model architecture of DiffWave is similar to that of WaveNet \cite{oord2016wavenet}. Without an autoregressive generation constraint, the dilated convolution is replaced with a bidirectional dilated convolution (Bi-DilConv). The non-autoregressive generation property of DiffWave yields a major advantage over WaveNet in that the generation speed is much faster. The network comprises a stack of $N$ residual layers with residual channel $C$. These layers were grouped into $m$ blocks, and each block had $n = \frac{N}{m}$ layers. The kernel size of Bi-DilConv is 3, and the dilation is doubled at each layer within each block as $[1,2,4,\cdots,2^{n-1}]$. Each of the residual layers has a skip connection to the output, which is the same as that used in Wavenet.

\begin{figure}[tp]
 \centering
 \includegraphics[width=\linewidth]{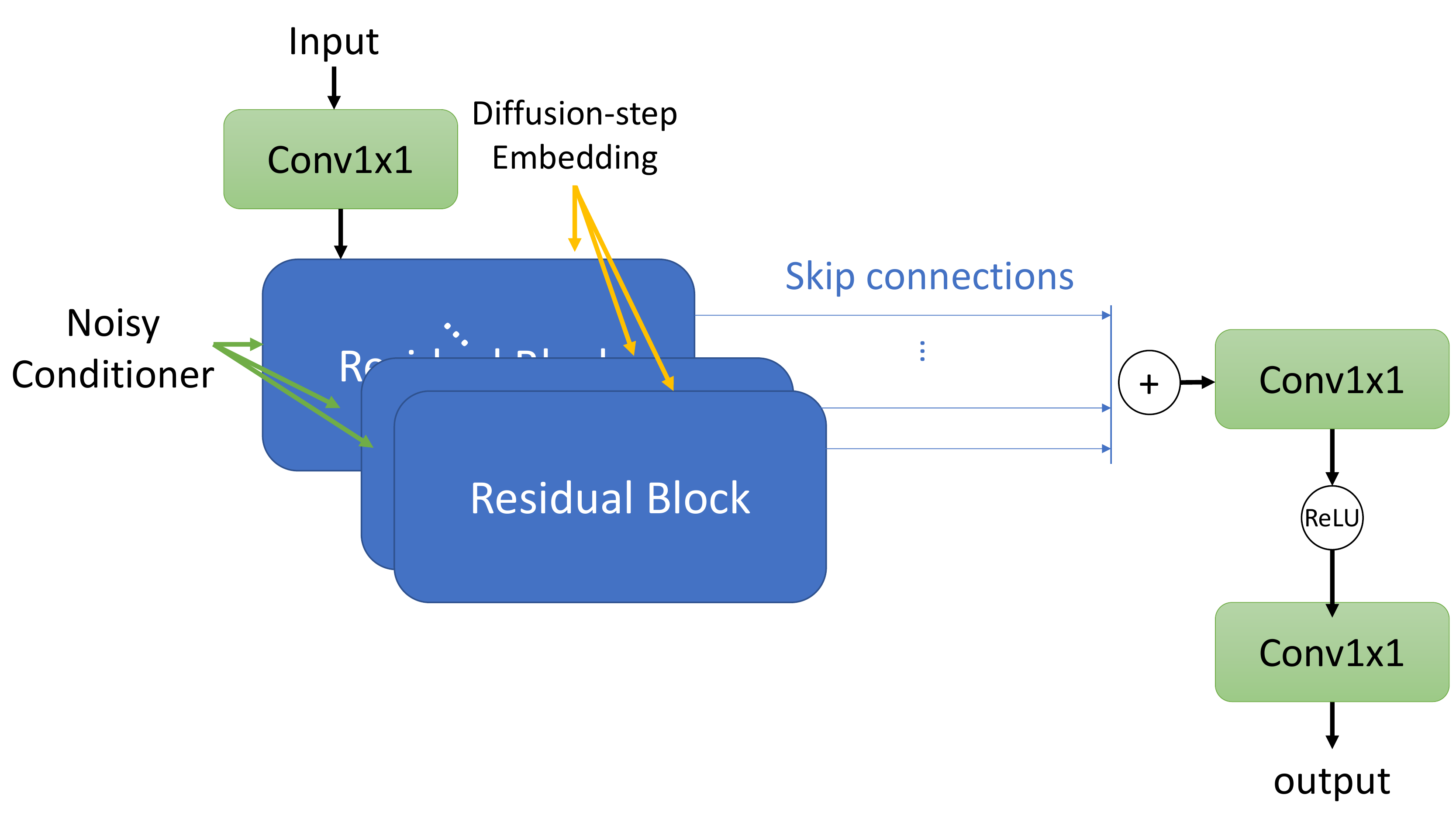}
 \caption{The architecture of the proposed DiffuSE model} 
 \label{fig:SE model_1}
\end{figure}

\subsubsection{DiffuSE Architecture}
Figure \ref{fig:SE model_1} shows the model structure of the DiffuSE. As Diffwave, the conditioner in DiffuSE aims to keep the output signal similar to the target speech signal, enabling $\epsilon_\theta(x_t,t)$ to separate the noise and clean speech from the mixed data. Thus, we replace the input of the conditioner from clean Mel-spectral features to noisy spectral features. We set the parameter of DiffuSE, $\epsilon_\theta : \mathbb{R}^L \times \mathbb{N} \rightarrow \mathbb{R}^{L}$, to be similar to those used in the DiffWave model \cite{kong2020diffwave}.

\subsection{Pretraining with Clean Mel-spectral Conditioner}
To generate high-quality speech signals, we pretrained the DiffuSE model with the clean Mel-spectral features. In DiffWave, the conditional information is directly adopted from the clean speech, allowing the model $\epsilon_\theta(x_t,t)$ to separate the clean speech and noise from the mixed-signals. After pretraining, we changed the conditioner from clean Mel-spectral features to the noisy spectral features, reset the parameters in the conditioner encoder, and preserved other parameters for the SE training.

\subsection{Fast Sampling}
Given a trained model from Algorithm \ref{alg:training}, the authors in \cite{kong2020diffwave} discovered that the most effective denoising steps in sampling occur near $t=0$ and accordingly derived a fast sampling algorithm. The algorithm collapses the $T$-step in the diffusion process into $T_{\text{infer}}$-step in the reverse process with a proposed variance schedule. This motivates us to apply the fast sampling into DiffuSE to reduce the number of denoising steps. In addition, by changing $\mu_\theta^{\text{fast}}(x_t,t)$ and $\sigma_t^{\text{fast}}$ to $\hat{\mu}_\theta^{\text{fast}}(x_t,t)$ and $\hat{\sigma}_t^{\text{fast}}$ using Eq. \ref{supportive reverse eq1} and Eq. \ref{supportive reverse eq2}, respectively, the fast sampling schedule can be combined with the supportive reverse process.

\section{Experiments}
\label{sec:pagestyle}
\subsection{Data}
We evaluated the proposed DiffuSE on the VoiceBank-DEMAND dataset \cite{valentini2016investigating}. The dataset contains 30  speakers from the VoiceBank corpus \cite{veaux2013voice}, which was further divided into a training set and a testing set with 28 and 2 speakers, respectively. The training utterances were mixed with eight real-world noise samples from the DEMAND database \cite{thiemann2013diverse} and two artificial (babble and speech shaped) samples at SNR levels of 0, 5, 10, and 15 dB. The testing utterances were mixed with different noise samples, according to SNR values of 2.5, 7.5, 12.5, and 17.5 dB to form 824 utterances (0.6 h). Additionally, utterances from two speakers were used to form a validation set for model development, resulting in 8.6 h and 0.7 h of data for training and validation, respectively. All of the utterances were resampled to 16 kHz sampling rates.

\begin{algorithm}[tp]
  \caption{Fast Sampling}\label{fast sampling}
  \begin{algorithmic}
      \State {Sample $x_T{\sim}p_{\text{latent}} =N(0,I),$} 
      \For{$s=T_{\text{infer}},T_{\text{infer}}-1,\cdots,1$}
        \State {Compute $\mu^{\text{fast}}_{\theta}(x_s,s)$ and ${\sigma^{\text{fast}}_s}$}
        \State {Sample $x_{s-1}\sim p_{\theta}(x_{s-1}|x_s)=$}
        \State {$N(x_{s-1};\mu^{\text{fast}}_\theta(x_s,s),{\sigma^{\text{fast}}_s}^2I)$}
      \EndFor
      \State \textbf{return} $x_0$
  \end{algorithmic}
\end{algorithm}

\subsection{Model Setting and Training Strategy}
The DiffuSE model was constructed using 30 residual layers with three dilation cycles $[1,2,\cdots,512]$ and a kernel size of three. Based on the design of DiffWave in \cite{kong2020diffwave}, we set the number of diffusion steps and residual channels as $[T,C]\in{[50,63],[200,128]}$ for Base and Large DiffuSE, respectively. The training noise schedule was linearly spaced as $\beta_t \in [1\times10^{-4},0.05]$ for Base DiffuSE, and $\beta_t \in  [1\times10^{-4},0.02]$ for Large DiffuSE. The learning rate was $2 \times 10^{-4}$ for both pretraining (using clean Mel-spectrum) and fine-tuning the DiffuSE model. The dimension for the Mel-spectrum was 80, and the dimension of the noisy spectrum was 513 for the same window size of 1024 with 256 shifts. The $\gamma_t$ parameter in the supportive reverse process 
was set to $\gamma_t = \frac{\sigma_t}{\sqrt{\bar{\alpha}_{t-1}}}$ for $t$ larger than 1, and $\gamma_1$ was set to 0.2 as the combination ratio of noisy signal to the enhanced output. During pretraining, we followed the instructions in \cite{kong2020diffwave}, where the vocoder model was trained for one million iterations, and the large model for three hundred thousand iterations for better initialization. In the training of the SE model, we trained the model for 300 thousand iterations for Base DiffuSE and 700 thousand iterations for Large DiffuSE. The batch size was 16 for Base DiffuSE and 15 for Large DiffuSE because of resource limitations. Both pretraining and fine-tuning DiffuSE are based on an early stopping scheme.


\subsection{Evaluation Metrics}
We report the standardized evaluation metrics for performance comparison, including perceptual evaluation of speech quality (PESQ) \cite{rix2001perceptual}, (the wide-band version in ITU-T P.862.2), prediction of the signal distortion (CSIG), prediction of the background intrusiveness (CBAK), and prediction of the overall speech quality (COVL) \cite{hu2007evaluation}. Higher scores indicated better SE performance for all of evaluation scores. 



\begin{table}[]
\centering
\caption{Evaluation results of (a) Base DiffuSE model and (b) Large DiffuSE model; both DiffuSE models adopted the original reverse process (RP) and the supportive reverse process (SRP). From ``RP", we further implemented ``RP-$N_{in}$" by replacing the Gaussian noise to noisy signal, and ``RP-$N_{out}$" by adding noisy signal at the generated output. ``RP-$N_{in+out}$" is a combination of ``RP-$N_{in}$" and ``RP-$N_{out}$". The results of the fast and full sampling schedules are listed as ``Fast" and ``Full", respectively. The results of the original noisy speech (denoted as ``Noisy") are also listed for comparison.}
\label{tab:DiffuSE Evaluation results}
\subfigure[Evaluation results of the Base DiffuSE model.]{
\begin{tabular}{cccccc}
\hline
Base DiffuSE & Schedule & PESQ & CSIG & CBAK & COVL \\ \hline
\multirow{1}{*}{Noisy} &-& 1.97 & 3.35 & 2.44 & 2.63\\ \hline
\multirow{2}{*}{RP} & Fast & 1.96 & 3.13 & 2.22 & 2.52 \\
 & Full & 1.97 & 3.21 & 2.22 & 2.57 \\ \hline
 \multirow{2}{*}{RP-$N_{in}$} & Fast & 2.07 & 3.21 & 2.57 & 2.62 \\
 & Full & 2.05 & 3.27 & 2.48 & 2.64 \\ \hline
\multirow{2}{*}{RP-$N_{out}$} & Fast & 2.05 & 3.31 & 2.21 & 2.64 \\
 & Full & 2.12 & 3.38 & 2.25 & 2.72 \\ \hline
\multirow{2}{*}{RP-$N_{in+out}$} & Fast & 2.29 & 3.47 & 2.67 & 2.85 \\
 & Full & 2.31 & 3.51 & 2.61 & 2.88 \\ \hline
\multirow{2}{*}{SRP} & Fast & \textbf{2.41} & \textbf{3.61} & \textbf{2.81} & \textbf{2.99} \\
 & Full & 2.38 & 3.60 & 2.79 & 2.97 \\ 
\hline
\label{tab:Base DiffuSE results}
\end{tabular}
}


\subfigure[Evaluation results of the Large DiffuSE model.]{
\begin{tabular}{cccccc}
\hline
Large DiffuSE & Schedule & PESQ & CSIG & CBAK & COVL \\ 
\hline
\multirow{1}{*}{Noisy} &-& 1.97 & 3.35 & 2.44 & 2.63\\ \hline
\multirow{2}{*}{RP} & Fast & 2.09 & 3.29 & 2.31 & 2.67 \\
 & Full & 2.16 & 3.39 & 2.31 & 2.75 \\ \hline
\multirow{2}{*}{RP-$N_{in}$} & Fast & 2.18 & 3.35 & 2.60 & 2.74 \\
 & Full & 2.20 & 3.42 & 2.48 & 2.78 \\ \hline
\multirow{2}{*}{RP-$N_{out}$} & Fast & 2.16 & 3.42 & 2.30 & 2.76 \\
 & Full & 2.17 & 3.45 & 2.29 & 2.78 \\ \hline
\multirow{2}{*}{RP-$N_{in+out}$} & Fast & 2.37 & 3.56 & 2.69 & 2.94 \\
 & Full & 2.33 & 3.55 & 2.56 & 2.91 \\ \hline
 \multirow{2}{*}{SRP} & Fast & \textbf{2.43} & \textbf{3.63} & \textbf{2.81} & \textbf{3.01} \\
 & Full & 2.39 & 3.63 & 2.75 & 2.99 \\ \hline
\label{tab:Large DiffuSE results}
\end{tabular}
}
\end{table}


\begin{table*}[htp]
\centering
\caption{Evaluation results of DiffuSE with comparative time-domain generative SE models. DiffuSE with the Base and Large models are denoted as DiffuSE(Base) and DiffuSE(Large), respectively. All of the metric scores for the comparative methods are taken from their source papers.}
\label{tab:DiffuSE results}
\begin{tabular}{p{3cm}p{2.5cm}p{2.5cm}p{2.5cm}p{2.5cm}p{2.5cm}}
 \hline	
Method & PESQ & CSIG & CBAK & COVL \\
 \hline	
Noisy & 1.97 & 3.35 & 2.44 & 2.63 \\
SEGAN & 2.16 & 3.48 & 2.94 & 2.80 \\
DSEGAN & 2.39 & 3.46 & 3.11 & 2.90 \\
SE-Flow & 2.28 & 3.70 & 3.03 & 2.97 \\
 \hline	
 \hline	
DiffuSE(Base) & 2.41 & 3.61  & 2.81  & 2.99\\
DiffuSE(Large) & 2.43 & 3.63  & 2.81  & 3.01\\
\end{tabular}
\end{table*}

\section{Experimental Results}
In this section, we first present the DiffuSE results with the  original reverse process and the proposed supportive reverse process. Next, we compare DiffuSE with other state-of-the-art (SOTA) time-domain generative SE models. Finally, we justify the effectiveness of DiffuSE by visually analyzing the spectrogram and waveform plots of the enhanced signals. 

\subsection{Supportive Reverse Process Results}
In the supportive reverse process, we adopted two sampling schedules, namely a fast sampling schedule and a full sampling schedule. For the fast sampling schedule, the variance schedules were $[0.0001,0.001,0.01,0.05,0.2,0.5]$ for Base DiffuSE and $[0.0001,0.001,0.01,0.05,0.2,0.7]$ for Large DiffuSE, as suggested in \cite{kong2020diffwave}. The full sampling schedule used the same $\beta_t$ as that used in the diffusion process. 

Tables \ref{tab:DiffuSE Evaluation results} (a) and (b) list the results of the Base DiffuSE model and the Large DiffuSE model, respectively. In the tables, the results of DiffuSE using the original reverse process and the supportive reverse processes are denoted as ``RP" and ``SRP," respectively. The table reports the results of both fast and full sampling schedules. To investigate the effectiveness of the supportive reverse process, we further tested performance by including noisy speech signal at the input, output, and both input and output of the DiffuSE model with the original reverse process; the results are denoted by ``RP-$N_{in}$," ``RP-$N_{out}$," and ``RP-$N_{in+out}$," respectively, in Table \ref{tab:DiffuSE Evaluation results}. When adding noisy speech at the input, we directly replaced the Gaussian noise with a noisy speech signal. When adding the noisy speech at the output, the final enhanced speech is a weighing average of the enhanced speech (80\%) and the noisy speech signal (20\%).

From Table \ref{tab:DiffuSE Evaluation results} (a), we first note that, except for RP, all of the DiffuSE setups achieved improved performance over ``Noisy" with a notable margin (for both fast and full sampling schedules).
Next, we observe that ``RP-$N_{in}$," ``RP-$N_{out}$," and ``RP-$N_{in+out}$" outperform ``RP," showing that including the noisy speech at the input and output can enable the original reverse process to attain better enhancement performance. 
Finally, we note that ``SRP" outperforms ``RP," ``RP-$N_{in}$," ``RP-$N_{out}$," and ``RP-$N_{in+out}$" for both fast and full sampling schedules, confirming the effectiveness of the proposed supportive reverse process for DiffuSE. 

Next, from Table \ref{tab:DiffuSE Evaluation results} (b), we observe that the results of the Large DiffuSE model present  trends similar to those of the Based DiffuSE model (shown in Table \ref{tab:DiffuSE Evaluation results} (a)). All of the DiffuSE setups provided improved performance over ``Noisy," and ``SRP" achieved the best performance among the DiffuSE setups. When comparing Tables \ref{tab:DiffuSE Evaluation results} (a) and (b), the Large DiffuSE model yielded better enhancement results than the Base DiffuSE model, revealing that a more complex DiffuSE model can provide better enhancement results.

From Tables \ref{tab:DiffuSE Evaluation results} (a) and (b), we notice that for ``RP" and ``RP-$N_{out}$," the full sampling schedule provided better results than the fast sampling schedule, which is consistent with the findings reported in DiffWave \cite{kong2020diffwave}. In contrast, for ``RP-$N_{in}$," ``RP-$N_{in+out}$," and ``SRP," the fast sampling schedule yielded better results than the full sampling schedule. A possible reason is that the noisy speech signal is a combination of clean speech and noise signals and presents clearly different properties from the pure Gaussian noise. Therefore, when including noisy speech in the input, it is more suitable to apply a fast sampling schedule than the full sampling schedule. 

\begin{figure}[htbp]
\centering
\subfigure[Clean]{
\includegraphics[width=4.1cm]{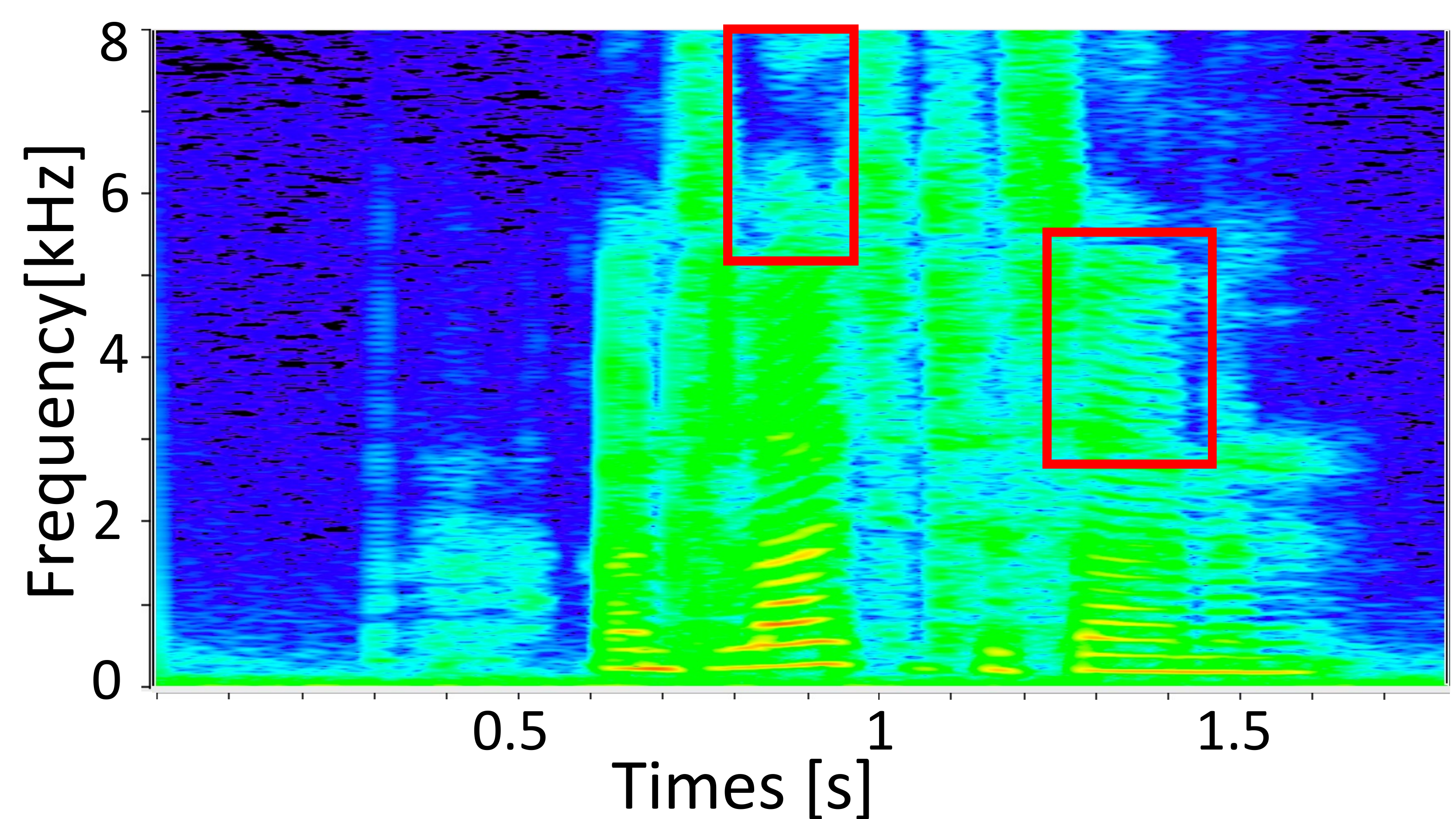}
}
\subfigure[Noisy]{
\includegraphics[width=4.1cm]{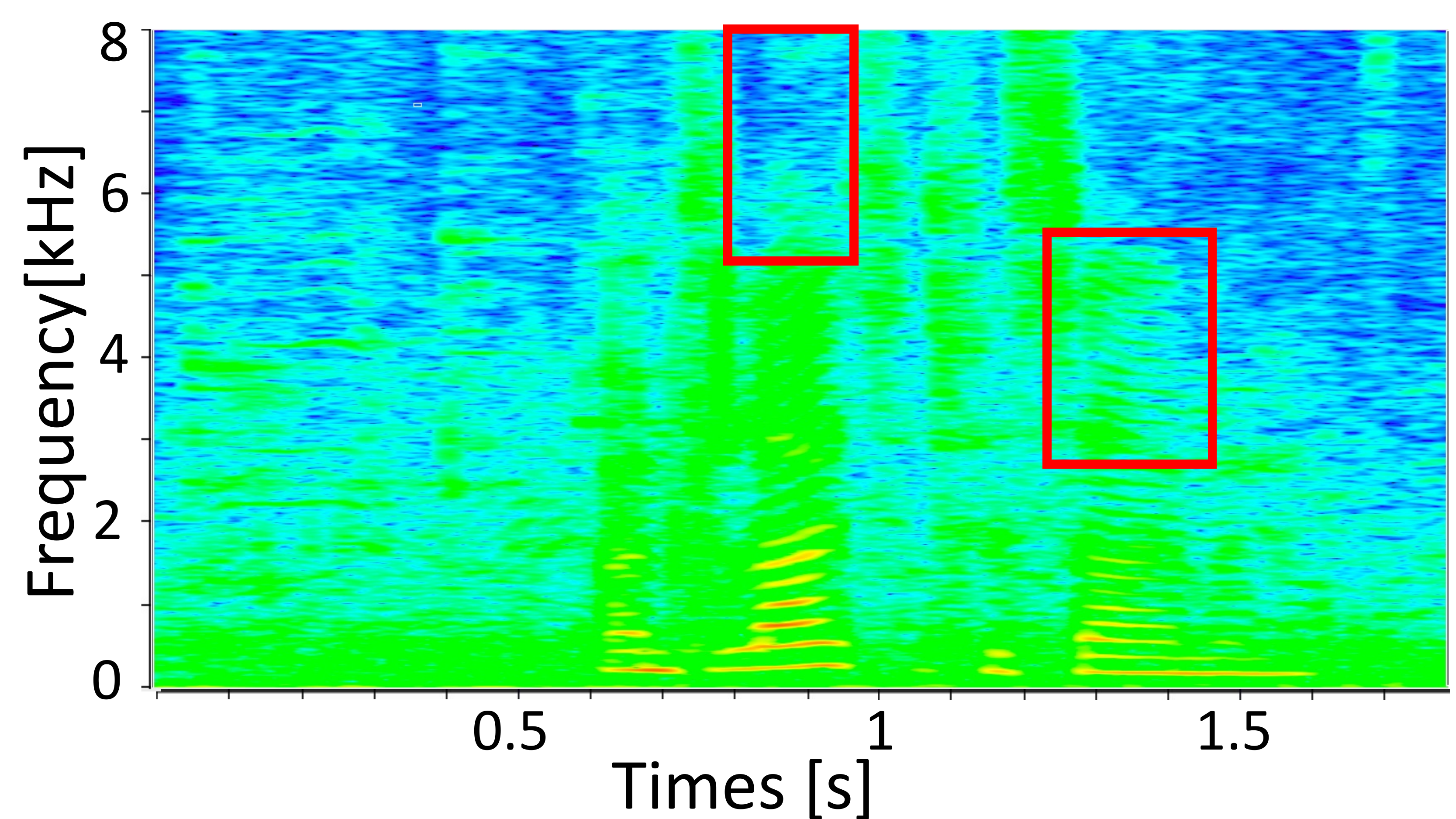}
}
\quad
\subfigure[DiffuSE+RP]{
\includegraphics[width=4.1cm]{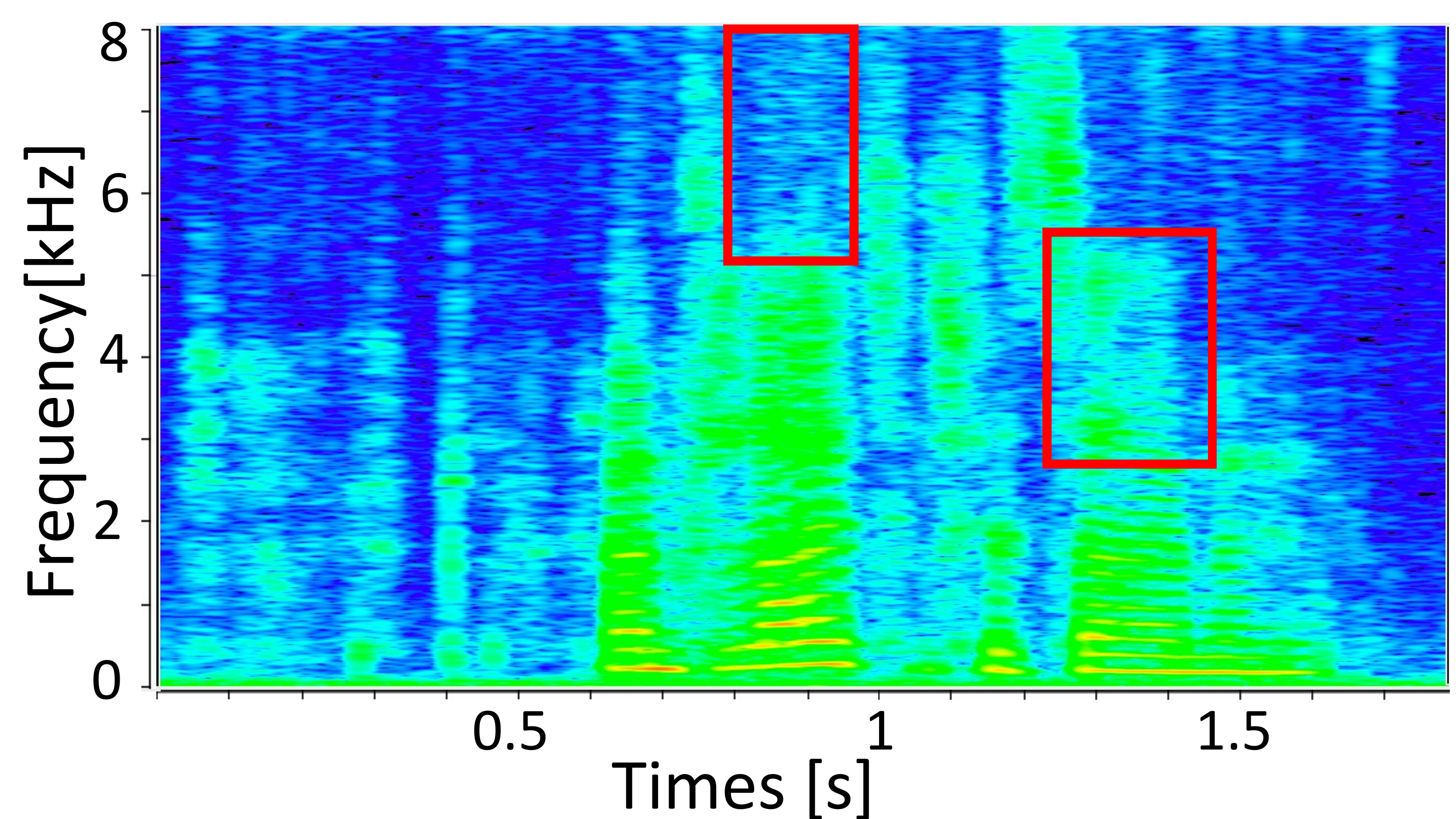}
}
\subfigure[DiffuSE+SRP]{
\includegraphics[width=4.1cm]{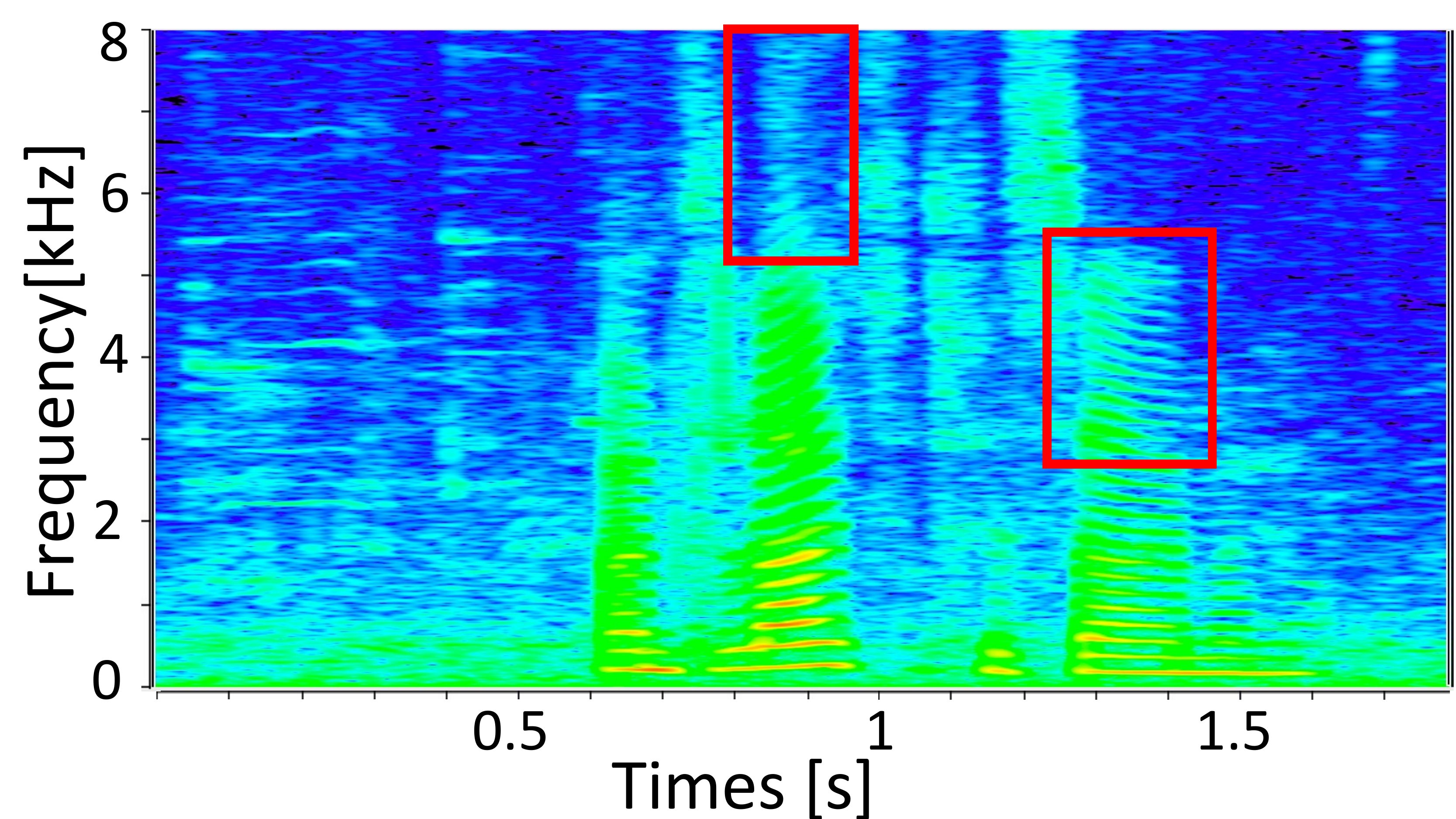}
}

\caption{Spectrogram plots of (a) Clean speech, (b) Noisy signal, (c) Enhanced speech by DiffuSE with the original reverse process (DiffuSE+RP) (d) Enhanced speech by DiffuSE with the supportive reverse process (DiffuSE+SRP).}
\label{fig:spectrogram}
\end{figure}
\begin{figure}[htbp]
\centering
\subfigure[Clean]{
\includegraphics[width=4.1cm]{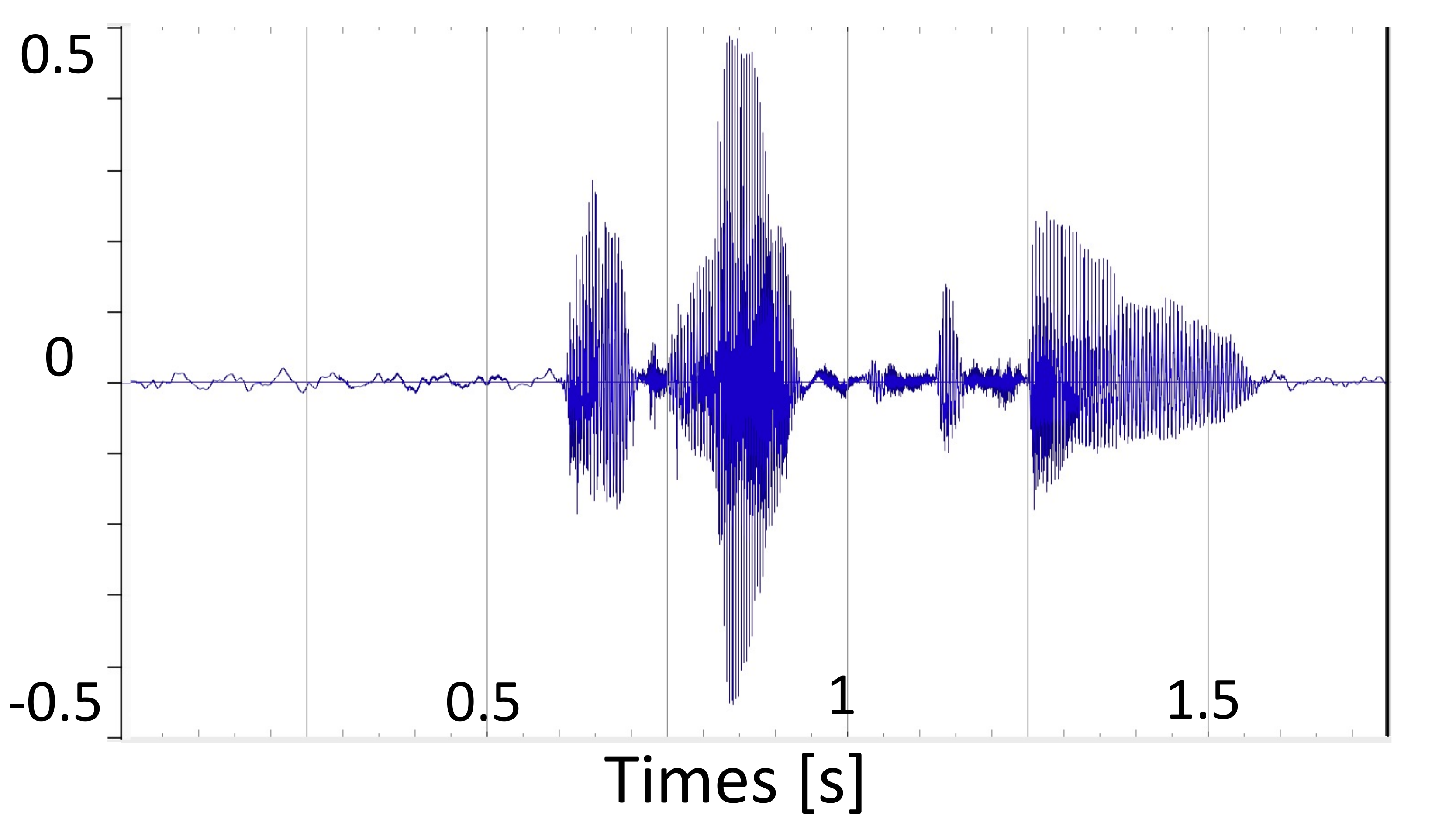}
}
\subfigure[Noisy]{
\includegraphics[width=4.1cm]{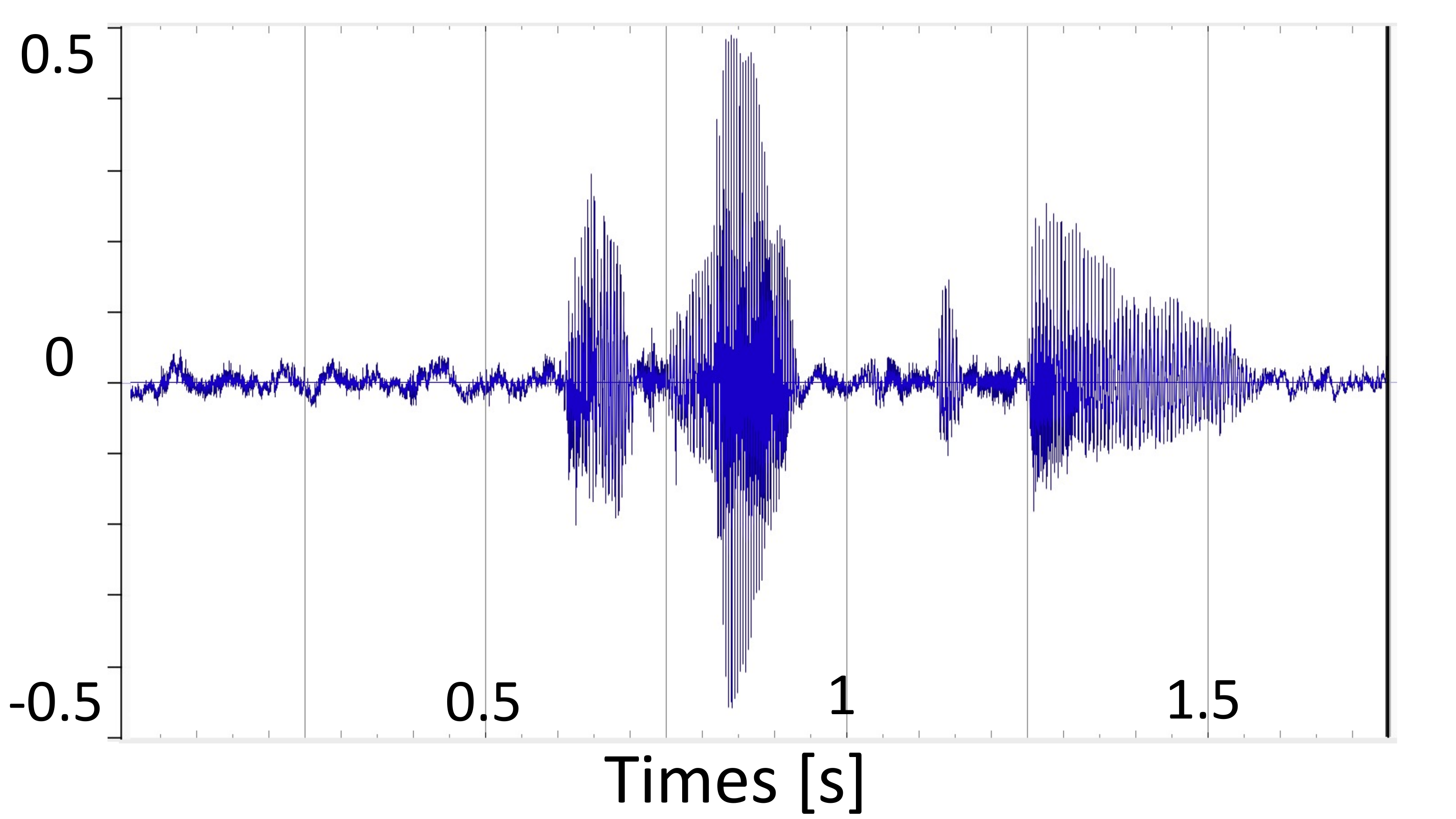}
}
\quad
\subfigure[DiffuSE+RP]{
\includegraphics[width=4.1cm]{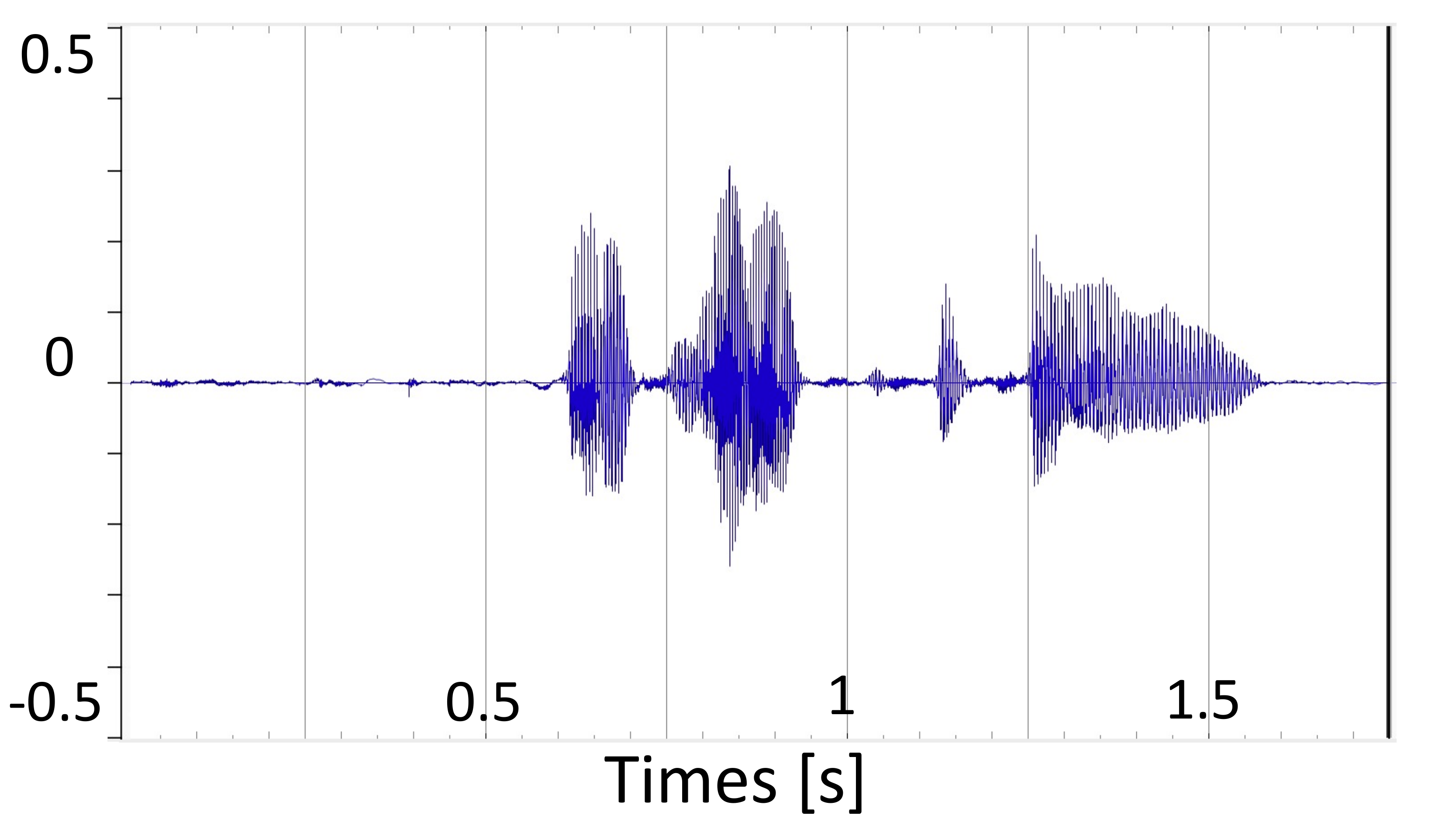}
}
\subfigure[DiffuSE+SRP]{
\includegraphics[width=4.1cm]{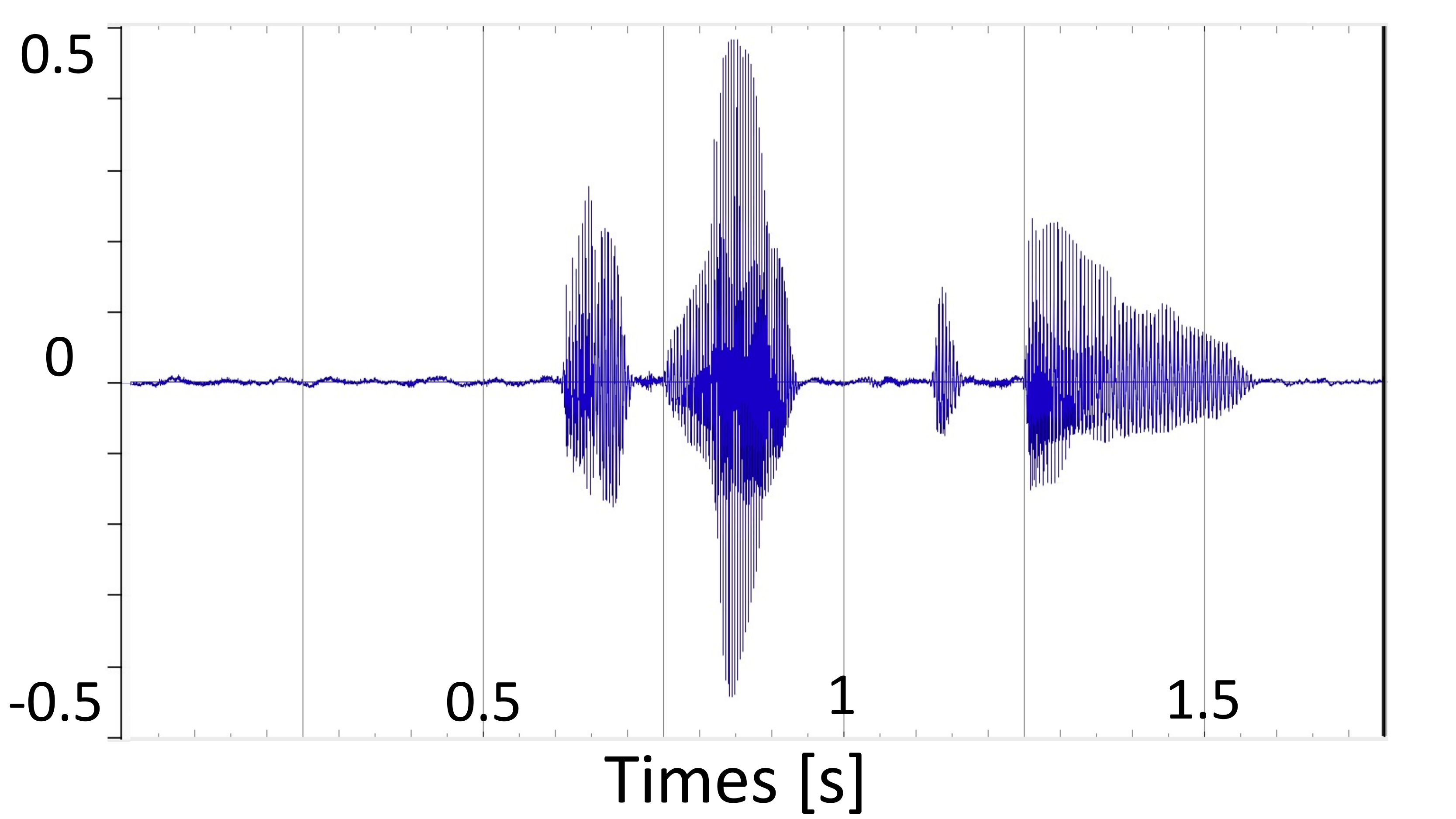}
}

\caption{Waveform plots of (a) Clean speech, (b) Noisy signal, (c) Enhanced speech by DiffuSE with the original reverse process (DiffuSE+RP) (d) Enhanced speech by DiffuSE with the supportive reverse process (DiffuSE+SRP).}
\label{fig:waveform}
\end{figure}

In addition to quantitative evaluations, we present spectrogram and waveform plots to qualitatively analyze the enhanced speech signals obtained from the DiffuSE models. Figures \ref{fig:spectrogram} and \ref{fig:waveform}, respectively, show the spectrogram and waveform plots of (a) clean, (b) noisy, (c) enhanced speech using DiffuSE with the original reverse process (denoted as DiffuSE+RP), and (d) enhanced speech using DiffuSE with the supportive reverse process (detonated as DiffuSE+SRP). From Figure \ref{fig:spectrogram}, we first note that both of the original and supportive reverse processes can effectively remove noise components from a noisy spectrogram. Next, we observe notable speech distortions in (c) DiffuSE+RP, especially in the high-frequency regions (marked with red rectangles). For (d) DiffuSE+SRP, although some noise components remained, the speech structures were better preserved as compared to (c) DiffuSE+RP. From Figure \ref{fig:waveform}, the waveform plots present similar trends to the spectrogram plots: the waveform of (d) DiffuSE+SRP preserves speech structures better than that of (c) DiffuSE+RP (please compare the two waveforms around 0.8 and 1.3 (s)). The observations in Figures \ref{fig:spectrogram} and \ref{fig:waveform} better explain the results obtained using the supportive reverse process over the original reverse process, as reported in Table \ref{tab:DiffuSE Evaluation results}. The samples of the DiffuSE-enhanced signals can be found online\footnote{\label{note1}https://github.com/neillu23/DiffuSE}.

\subsection{Comparing DiffuSE with Related SE Methods}
The proposed DiffuSE model is a time-domain generative SE model. For comparison, we selected three SOTA baselines that are also based on time-domain generative SE models, namely SEGAN \cite{pascual2017segan}, SE-Flow \cite{strauss2021flow}, and improved deep SEGAN (DSEGAN) \cite{phan2020improving}. 
The experimental results of the three comparative SE methods are presented in Table \ref{tab:DiffuSE results}. The results of the DiffuSE with the supportive reverse process are also listed, where DiffuSE(Base) and DiffuSE(Large) denote the results of using the base and large models, respectively. Compared with the three baselines, the PESQ scores of DiffuSE(Base) and DiffuSE(Large) are 2.41 and 2.43, respectively, both of which are much higher than those obtained from the comparative methods. The CSIG scores of DiffuSE(Base) and DiffuSE(Large) are 3.61 and 3.63, respectively, again notably higher than those achieved by SEGAN and DSEGAN. The results confirm that the proposed DiffuSE method provides a competitive performance against SOTA generative SE models.

\section{Conclusions}
In this study, we have proposed DiffuSE, the first diffusion probabilistic model-based SE method. To enable an efficient sampling procedure, we proposed modifying the reverse equation to a supportive reverse process, specially designed for the SE task. Experimental results show that the supportive reverse process can improve the quality of the generated speech with few steps to obtain better performance than that of the full reverse process. The results also show that DiffuSE achieves SE performance comparable to that of other 
SOTA time-domain generative SE models. The results of DiffuSE are reproducible and the code of DiffuSE will be released online$^1$. We believe that the results will shed light on further extensions of using the diffusion probabilistic model for the SE task. In future work, we will further improve the DiffuSE model through different network structures.

\section{Acknowledgement}
This work was supported in part by the grants AS-GC-109-05 and AS-CDA-106-M04 and we would like to thank Alexander Richard at Facebook for his valuable comments about this work.










\bibliographystyle{IEEEbib}
\bibliography{strings,refs}
\end{document}